# Temperature dependence of activation energy of the conductivity of manganites in paramagnetic phase

E.A.Neifeld,[1] V.E.Arkhipov, N.A.Ugryumova, A.V.Korolyov, Ya.M. Mukovsky*

Institute of Metal Physics, Ural Division of RAS, Kovalevskaya str. 18, 620219 Ekaterinburg Russia

*Moscow State Institute of Steel and Alloys

Russia

The experimental results of studying the resistivity and magnetic susceptibility are presented for the single crystal $La_{1-x}Sr_xMnO_3$ ($x = 0.15; 0.175$) and $CaMnO_{3-\delta}$. We revealed the temperature dependence of the magnetic component of activation energy of hopping conductivity $(\varepsilon_m)$ in paramagnetic phase: $\varepsilon_m = \varepsilon_m^0 \pm a/T$, where the «+» sign corresponds to antiferromagnetic manganite $CaMnO_{3-\delta}$, and the «-» sign to ferromagnetic manganites $La_{1-x}Sr_xMnO_3$. It has been shown that for antiferromagnetic $CaMnO_{3-\delta}$ that temperature dependence of activation energy has a maximum coinciding with a maximum of magnetic susceptibility. For ferromagnetic sample $La_{0.85}Sr_{0.15}MnO_3$ the minimum of activation energy and maximum of magnetic susceptibility also coincide. The results obtained are explained on the basic of the model proposed in Gor'kov's works.

To understand the nature of colossal magnetoresistance effect observed in manganites near the Curie temperature ($T_C$) it is necessary to analyze the mechanism of the conductivity in paramagnetic phase transfer is realized i.e. at $T > T_C$. As is well known in this temperature range the charge, transported by hopping of polarons with a small radius [1]. In this case the temperature dependence of the resistivity has the form:

$$\rho \sim T \exp(\varepsilon_a / kT),$$

where $\varepsilon_a$ is the activation energy. In majority of experimental works the activation energy is considered to be a constant, i.e. independent of temperature and is determined by linear approximation of experimental dependence of $\rho(T)$ in coordinates $\ln(\rho / T)$; $T^{-1}$.

---

[1] Neifeid@imp.uran.ru

However, more detailed analysis allowing us to reveal the temperature dependence of $\varepsilon_a(T)$ consists in determining the local activation energy

$$\varepsilon_a = d(\ln(\rho/T))/d(T^{-1}). \quad (1)$$

In this work we studied the activation conductivity in paramagnetic phase of ferromagnetic $La_{1-x}Sr_xMnO_3$ ($x = 0.15; 0.175$) and of antiferromagnetic $CaMnO_{3-\delta}$. One of ferromagnetic samples ($x = 0.15$) at $T > T_C$ showed the activation conductivity and another ($x = 0.175$) had metallic one.

Single crystals were grown by the floating zone method ($La_{1-x}Sr_xMnO_3$) and zone melting ($CaMnO_{3-\delta}$). Electrical resistivity was measured by the four-probe technique. The dynamic magnetic susceptibility was measured with SQUID magnetometer MPMS-5XL of Quantum Design with the alternating magnetic field amplitude of 4Oe.

More detailed characteristics of the $La_{1-x}Sr_xMnO_3$ samples are given in [2] ($x = 0.15$) and [3] ($x = 0.175$); the characteristics of the $CaMnO_{3-\delta}$ sample can be found in the work [4] the authors of which by the neutron-diffraction studies have shown that $\delta = 0.25$.

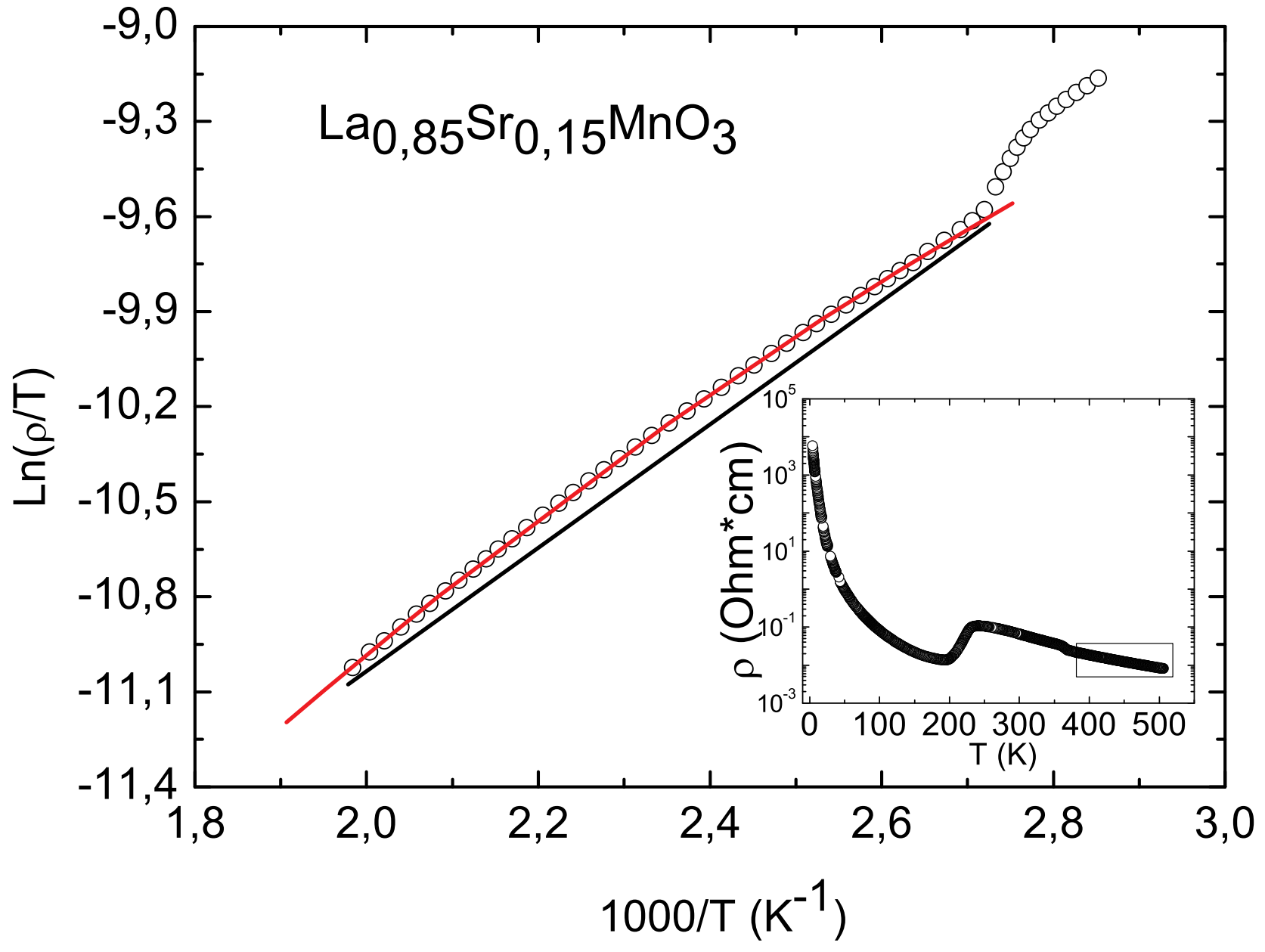

Fig.1. Logarithm of resistivity vs. the inverse temperature for $La_{0.85}Sr_{0.15}MnO_3$ in the temperature range of 365-500K. Inset: resistivity vs.temperature for this sample in the range of 4.2-500K; rectangular frame marks the portion of the curve presented in the principal figure.

The inset of Fig. 1 shows general view of the dependence $\rho(T)$ for $La_{0.85}Sr_{0.15}MnO_3$; the frame picks out the region which in the principal figure is depicted in coordinates [$\ln(\rho/T)$; $T^{-1}$]. The straight line drawn near the experimental values demonstrates that linear approximation of experimental data is rather rough. The curve passing through the experimental points is a result of fitting the data obtained with a quadratic polynomial. (For visual demonstration a number of experimental points in this plot was approximately diminished by a factor of six). The results for antiferromagnetic $CaMnO_{3-\delta}$ are presented in Fig. 2 in a similar form. The experimental dependence of $\ln(\rho/T)$ on $T^{-1}$ is also well described by a quadratic polynomial but with the unlike curvature sign.

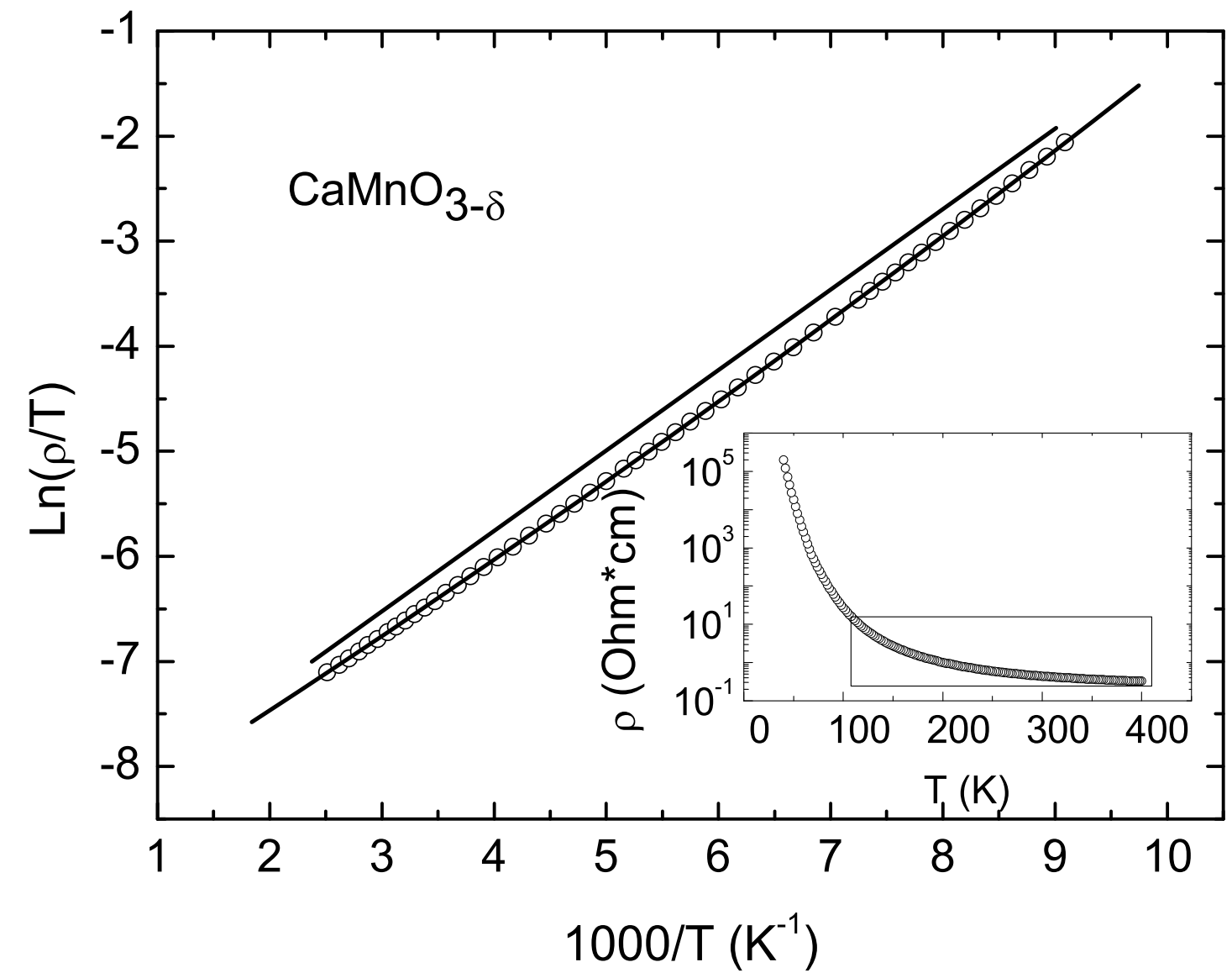

Fig. 2. Logarithm of resistivity vs. the inverse temperature for $CaMnO_{3-\delta}$ in the temperature range of 110-400K. Inset: resistivity vs. temperature for this sample in the range of 40-400K; frame marks the portion of the curve presented in the principal figure.

It should be noted that we studied previously [2] the temperature dependence of activation energy for $La_{0.85}Sr_{0.15}MnO_3$ in the temperature range of 4,2 – 400K. However, the high-temperature interval in which the dependence $\rho(T)$ was analyzed had been small. It is bounded by the structural transition (~350K) and magnetic one ($T_C$=220K). As far as the influence of magnetic transition on the activation energy magnitude manifests itself at temperatures considerably higher than $T_C$, we have wrongly concluded that in the paramagnetic phase the variable range hopping (VRH) conductivity takes place. In the present work the electrical resistivity was analyzed at temperatures significantly higher than $T_C$ and even above the temperature of structural transition, but no indications of VRH conductivity (linear dependence $\rho(T^{1/4})$) had been revealed.

Since the logarithmic plots of the resistivity depending on the inverse temperature are described by a quadratic polynomial (Figs. 1,2), it is obvious that local activation energy (1) determined from this plot will linearly depend on $T^{-1}$. For $La_{0.85}Sr_{0.15}MnO_3$ this dependence is described by the expression $\varepsilon_a = (349 - 78210/T(K))meV$ and for $CaMnO_{3-\delta}$ this expression has the form $\varepsilon_a = (57 + 1477/T(K))meV$. This means that in the paramagnetic phase in ferromagnetic manganite with lowering temperature the activation energy of hopping conductivity decreases by a value inversely proportional to temperature and in antiferromagnetic manganite it on the contrary increases. The temperature dependences of local activation energy is also shown for these samples are presented in Fig. 3, in which the temperature dependence of the activation energy for $La_{0.825}Sr_{0.175}MnO_3$ ($\varepsilon_a = (166 - 29296/T(K))meV$) which at $T > T_C$ falls into metallic side of the concentration transition. The values of $\varepsilon_a$ obtained by the linear approximation of experimental $\ln(\rho/T)$-vs-$T^{-1}$ plots are shown by dotted lines.

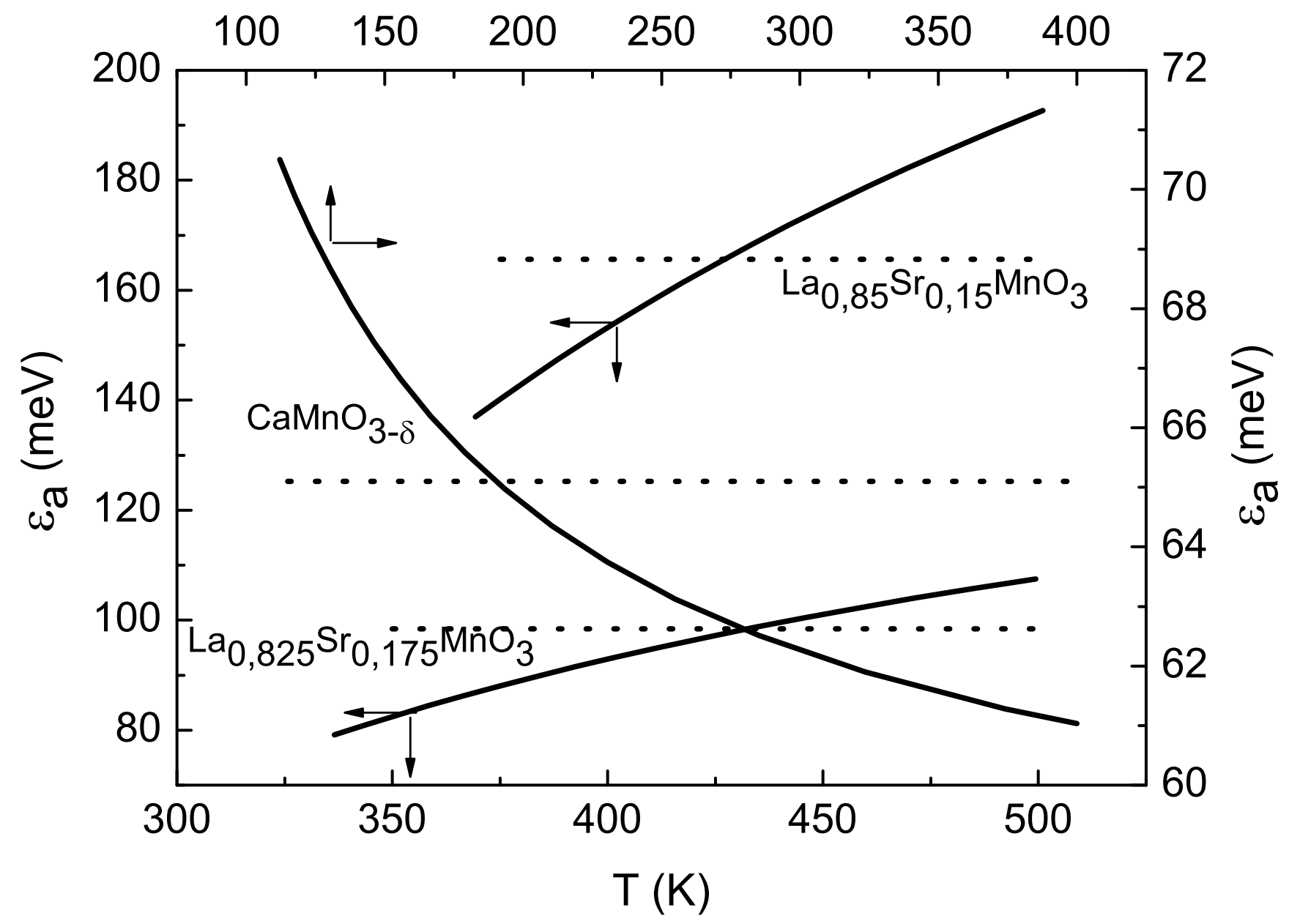

Fig.3. Temperature dependences of local activation energies for the indicated samples in the paramagnetic phase. Dotted lines show the activation energy values obtained by the linear approximation of corresponding portions of $\ln(\rho/T)$-vs-$T^{-1}$ curves.

It is known that in manganites the activation energy of hopping conductivity depends not only on Coulomb and lattice interactions but also on magnetic interactions [5]. We believe that experimental linear dependence of the activation energy on the inverse temperature is related to the temperature dependence of magnetic component ($\varepsilon_m$)

$$\varepsilon_m = \varepsilon_m^0 \pm a/T \quad , \qquad (2)$$

where $\varepsilon_m^0$ is the magnetic component of the activation energy at $T \to \infty$, $a$ is a constant. The plus and minus signs correspond at low temperatures to manganites with antiferromagnetic and ferromagnetic orders, respectively. The same temperature dependence of the activation energy in the paramagnetic phase has been observed by us for other manganites $La_{1-x}Sr_xMnO_3 (x = 0.1; 0.2)$, $La_{1-x}Ba_xMnO_3 (x = 0.1; 0.15; 0.2, 0.25, 0.3)$, $\mathrm{Pr}_{1-x}Sr_xMnO_3 (x = 0.22; 0.24)$, $La_{0.6}Sr_{0.4}MnO_3$ in the temperature range in which magnetic susceptibility obeys the Curie-Weiss law.

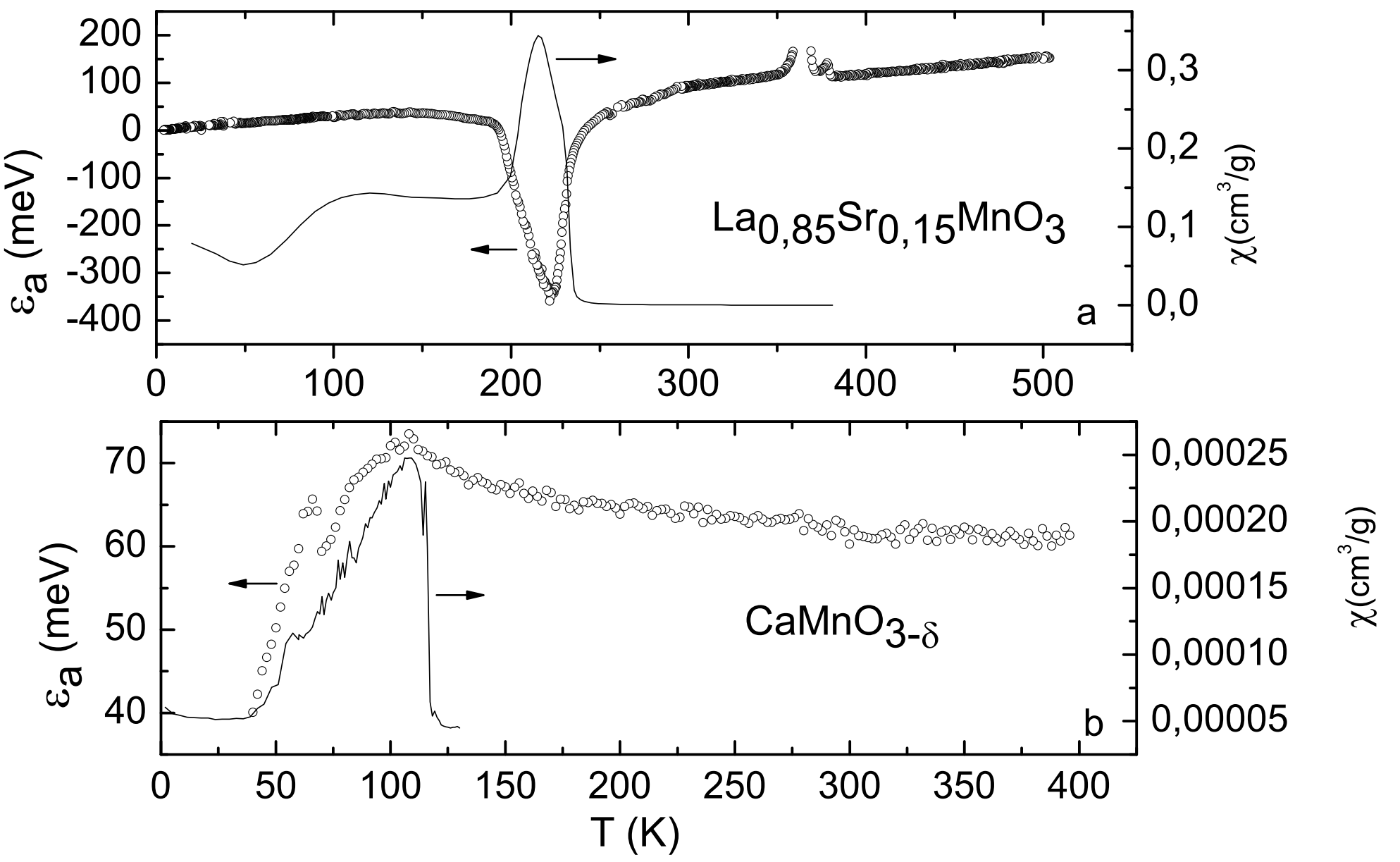

Fig.4. Temperature dependences of local activation energies and magnetic susceptibilities in the whole temperature range studied for $La_{0.85}Sr_{0.15}MnO_3$ (a) and $CaMnO_{3-\delta}$ (b).

The linear dependence of activation energy on the inverse temperature allows us to suppose that this dependence as well as the temperature dependence of magnetic susceptibility have the same origin. Therefore, in Fig. 4 we presented the temperature dependences of activation energy and magnetic susceptibility for $La_{0.85}Sr_{0.15}MnO_3$ and $CaMnO_{3-\delta}$ in more wide temperature range including the temperature of magnetic transition. Let us analyze at first the data obtained for $La_{0.85}Sr_{0.15}MnO_3$ (Fig.4a). Of course, here negative value of activation energy near $T_C$ have no physical meaning and are related to the procedure of determining $\varepsilon_a$. However, as far as the maximal value of conductivity (see the insert in Fig. 1) near $T_C$ is less than minimal metallic conductivity, one can deduce that the conductivity has the activation character in the whole temperature range. The temperature corresponding to a minimum of the real activation energy coincides with the minimum of $\varepsilon_a(T)$ dependence in Fig. 4a and coincides practically with the maximum of magnetic susceptibility. In the case of antiferromagnetic $CaMnO_{3-\delta}$ (Fig. 4b) the maximum of the temperature dependence of activation energy is observed at the same temperature as maximum of magnetic susceptibility. A peculiarity at T≅60K may be evidently related to the inhomogeneity of a sample.

We were able to explain the $\varepsilon_a(T)$ dependence observed in manganites in the framework of Gor'kovs model [6] in which phase segregation is considered as a dynamic state and assumed that a hole initiated by the impurity ion is not localized at any ion of Mn but is shared by all eight ions of Mn surrounding this impurity ion. Ferro- and antiferro-magnetocorrelated fluctuations can exist in such a media in the form of a “foggy” state. Their sizes are determined by the electroneutrality condition. Electrical conductivity arises in ferromagneto-correlated fluctuations as a result of double exchange. The degree of magnetic ordering and consequently magnetic component of activation energy of hopping conductivity depend on the ratio of the interaction energy of Mn ion magnetic moments and thermal energy *kT*. This explains the analogy of the temperature dependences of activation energy of hopping conductivity and magnetic susceptibility which value is determined by the ratio of the interaction energy of magnetic moments with a weak external magnetic field and thermal energy.

Thus, in the paramagnetic phase of manganites the temperature dependence of magnetic component of activation energy of hopping conductivity (2) is determined by the competition of ferro- and antiferromagnetic exchange interactions. In ferromagnetic manganites with decreasing temperature the activation energy decreases because of the increase of the ferromagnetic correlation degree of fluctuations, in other words, because of the decrease of the angle between magnetic moments of the Mn ions transferring a charge irrespective of whether the compound is a dielectric or a metal at $T < T_C$. In antiferromagnetic manganites the antiferromagnetic fluctuations interfering with the formation of ferromagnetic fluctuations predominate and with lowering temperature the activation energy rises. With temperature decreasing of ferromagnetic samples such a process is observed until spontaneous magnetization domains arise which give rise to the appearance of a "tail" in magnetic susceptibility [7] and to more sharp decrease of activation energy. In solid solutions these domains manifest themselves at the temperatures significantly higher that $T_C$. In the $La_{1-x}Sr_xMnO_3$ crystals studied a sharp decrease of the activation energy is observed at $T = 300K$ for $x = 0.15\,(T_c = 220K)$ and at $T = 340K$ for $x = 0.175\,(T_c = 280K)$. In the case of a sample with $x = 0.15$ this estimation is somewhat arbitrary because the sought for temperature falls into the interval between the structural and magnetic transitions and, as was mentioned above, the analysis of the dependence becomes unreliable. On the contrary, in antiferromagnetic $CaMnO_{3-\delta}$ the dependence (3) is practically valid down to the Néel temperature $(T_N = 116K)$.

In ferromagnetic phase for the manganites in dielectric region of the concentration metal-dielectric transition the activation energy decreases with decreasing temperature as the conductivity takes the VRH character [2]. We believe that in antiferromagnetic phase of $CaMnO_{3-\delta}$ the reason of decreasing the activation energy is probably the same but we could not make sure of this because of very large electrical resistivity of a sample at low temperatures.

The authors are indebted to A.M.Balbashov for placing at our disposal the $CaMnO_{3-\delta}$ crystal for study. The work was fulfilled in the program of RAS "New Materials and Structures" and supported by RFFI (grant 05-02-16303).